\documentclass[aps.pra,onecolumn,showpacs,preprintnumbers,amsmath,amssymb]{revtex4}

\usepackage[utf8]{inputenc}
\usepackage[english]{babel}
\usepackage{setspace}
\usepackage{booktabs}
\usepackage{multirow}
\usepackage{amssymb}

\usepackage{caption}
\usepackage{lipsum}
\usepackage{natbib}
\usepackage{graphicx}
\usepackage{braket}
\usepackage[figurename=Fig., justification=RaggedRight]{caption}
\usepackage{amsmath}
\usepackage{xcolor}
\usepackage{hyperref}

\begin{document} 

\title{Quantum simulation of fermionic systems using hybrid digital-analog quantum computing approach}

\author{N. M. Guseynov$^{1,2}$, W. V. Pogosov$^{1,3,4}$}
\affiliation{$^1$Dukhov Research Institute of Automatics (VNIIA), Moscow, Russia}
\affiliation{$^2$Moscow State University, Moscow, Russia}
\affiliation{$^3$Institute for Theoretical and Applied Electrodynamics, Russian Academy of Sciences, Moscow, Russia}
\affiliation{$^4$HSE University, Moscow, Russia}

\begin{abstract}
We consider a hybrid digital-analog quantum computing approach,
which allows implementing any quantum algorithm without standard
two-qubit gates. This approach is based on the always-on interaction between qubits, which can provide an alternative to such gates.
We show how digital-analog approach can be applied to simulate
the dynamics of fermionic systems, in particular, the Fermi-Hubbard
model, using fermionic SWAP network and refocusing technique.
We concentrate on the effects of connectivity topology, the spread of interaction constants as well as on errors of entangling operations.
We find that an optimal connectivity topology of qubits for the
digital-analog simulation of fermionic systems of arbitrary dimensionality is a chain for spinless fermions and a ladder for spin 1/2 particles. Such a simple connectivity topology makes digital-analog approach attractive for the simulation of quantum materials and molecules. 
\end{abstract}

\author{}
\maketitle
\date{\today }

\maketitle

\section{Introduction}

Quantum computing is a new computing paradigm, which is based on principles of quantum mechanics \cite{FEYNMAN}. Various physical platforms can be utilized to construct quantum computers: superconducting quantum circuits \cite{ex_sconduct}, linear optical quantum systems \cite{q_s_phot,q_c_nowadays}, trapped ions \cite{ions1,ions2}, neutral atoms \cite{neutral_atom}, etc. The advantage of quantum computing is that it allows one to tackle some problems that are hard to solve using conventional computing systems. Quantum machines can be used for mathematical problems, such as quantum search algorithm \cite{grover,q_search}, factoring integers by Shor's algorithm \cite{Shor,math_apply}, solving systems of linear equations \cite{PhysRevLett.103.150502}, etc. Another promising application of quantum computing is a quantum simulation, as proposed originally by R. Feynman \cite{FEYNMAN,VQE_molecules,Fermions,F_H,MAPPING_TO_QUBITS}. For example, the number of sampling points required for a classical simulation of the Fermi-Hubbard model at low temperatures using Monte Carlo methods rises exponentially as the system volume grows, but quantum computing can provide tools to determine its zero-temperature phase diagram. Note that the Fermi-Hubbard model plays a central role in theoretical studies of strongly correlated materials \cite{Hubbard1,Hubbard2,Hubbard3}. Particularly, this model is of importance to understand their magnetic and superconducting properties as well as a complex interplay between them. Therefore, the development of methods of quantum simulation for fermionic systems is an extremely important task.

The ability of quantum computers to outperform classical supercomputers has recently been demonstrated for some special tasks. This regime of functioning of quantum computers corresponds to the so-called quantum supremacy \cite{q_s_phot,q_c_nowadays,Q_supremacy1,Q_supremacy2}. Conventional quantum computers operate in accordance with the gate model of quantum computation which is based on a universal set of quantum gates. The minimal set of required gates \cite{Q_evidence} includes arbitrary single-qubit gates and two-qubit operations such as controlled-NOT (CNOT) or some equivalent gate. Such quantum devices are usually referred to as digital quantum computers. However, control of all relevant degrees of freedom becomes much more difficult when the system is scaled up. Quantum error correction codes \cite{errors} in combination with optimization protocols \cite{VQE_molecules,optimise1,optimise2} constitute a promising approach, but their effective implementation on modern quantum devices is challenging due to the relatively small number of qubits as well as existing gates errors. As a result of hardware imperfections, the Fermi-Hubbard model was simulated using digital quantum computers only for small systems with limited numbers of fermionic lattice sites \cite{Obs_sep_dyn}. Note that the main contribution to the gates error rate is due to the two-qubit gates. Thereby, it is interesting to consider alternative approaches, which are not based on the execution of the standard two-qubit gates.

One of the alternatives is associated with the idea of a hybrid digital-analog quantum computing \cite{35,36,37,Walter,d-a-f-h,ying2021floquet,PhysRevResearch.2.013012,QAOADA,DASC,Arrazola2016,Lamata2017,PRXQuantum.2.020328}. In the simplest realization of such a strategy, single-qubit gates, as well as individual readouts of qubits, must be accessible similarly to the case of digital quantum devices, but instead of conventional two-qubit gates quantum entanglement is created using always-on native interaction of qubits. This requires a certain trade-off between the fidelities of single-qubit gates diminished by the always-on interaction between qubits and fidelities of entangling operations \cite{Walter}. A more sophisticated approach is possible which requires switchable interaction between qubits that can be turned off during the execution of single-qubit gates. Particularly, we would like to mention a recent paper \cite{DASC}, which suggests a specialized superconducting circuit architecture for the digital-analog quantum computation and simulation. 

In this paper, we treat a native interaction of Ising type as a source of entangling operations and analyze the digital-analog implementation of quantum simulation of fermionic systems. Particularly, we address a method described in Ref. \cite{FSG} for digital quantum computers, which is based on the fermionic SWAP network. The fact that the interaction is always on implies that quantum evolution is governed by operators of pairwise interaction between qubits coupled according to the qubit network topology. Moreover, each qubit pair is, in general, characterized by its own coupling constant. Despite this fact, it turns out that using sequences of single-qubit gates interleaved with idling time intervals it is possible to compensate undesirable interactions and to engineer a needed global unitary transformation. The basic ideas behind this technique were developed in quantum computing studies based on nuclear magnetic resonance (NMR) \cite{refocusing1,refocusing2,refocusing3}. We incorporate these refocusing operations into the fermionic SWAP network. We then extend the concept of the refocusing operation to the situation, when the spread in the interaction constants is present. We also focus on the impact of different connectivity topologies of qubit networks and the effect of errors in the implementation of entangling operations.

This paper is organized as follows. We formulate the problem in Section II. Section III deals with the fermionic SWAP network
implemented using the refocusing technique. This Section, in particular, focuses on the impact of connectivity topology of
the qubit network. Section IV deals with the qubit networks having a spread in interaction constants. In Section V we discuss the effect of errors of entangling operations. We conclude in Section VI.

\section{Preliminaries}

Two-qubit operations between qubits $p$ and $q$ in a
digital-analog device correspond to the action of a quantum operator of the form $J_{pq}=e^{- i H_{int}^{p,q}t}$, where $H_{int}^{p,q}$ is the interaction Hamiltonian of two qubits. In the present paper, we consider Ising coupling between qubits, which is relevant for various physical implementations of quantum devices. For example, superconducting qubits can be coupled directly by $XX$ dipole-dipole interaction. Such coupling can be also mediated by some bosonic mode, such as a quantized photon field of the cavity coupled to both qubits. In the latter case, it is possible to eliminate perturbatively the boson mode from the effective description, provided the system is in a dispersive regime, which again results in the $XX$ interaction between the qubits. In Ref. \cite{Walter} $ZZ$-crosstalks between superconducting transmons-qubits were utilized as a source of native interaction used to implement a quantum simulation of spin models as well as the quantum Fourier transform. Ising coupling is also natural to some ion-trapped implementations of quantum computers \cite{ion}.

The evolution operator in the appropriate basis takes the form $J_{pq}=e^{- i Z_p Z_q t}$, where $t$ is a dimensionless time measured in terms of the interaction constant. This time is a free parameter for the entangling operation. Hereafter, $X_j$, $Y_j$, $Z_j$ refer to the Pauli operators acting in the space of $j$'th qubit. Of course, a direct application of operators $J_{pq}$ is possible only for those qubits which can interact according to the connectivity topology of the qubit network. The full interaction Hamiltonian, in this case, has the Ising-type form
\begin{eqnarray}
H_{int}=\hbar\sum_{p>q}\epsilon_{pq}Z_pZ_q,
\label{Hint}
\end{eqnarray}
where $\epsilon_{pq}$ is 1 or 0 depending on connectivity topology.  We hereafter denote $J(t)=\prod_{p,q} J_{pq}(t)$  a global evolution operator corresponding to the physical interaction between the qubits of the network according to its topology.

In this article, we discuss a method of quantum simulation of
fermionic systems, which uses a concept of the fermionic SWAP network. This
smart method was developed in Ref \cite{FSG}. It allows one to circumvent
difficulties associated with the mismatch between the fermion statistics of the simulated system
and paulion statistics relevant for qubits. Our aim is to adapt this
approach to the digital-analog quantum computing strategy based on
always-on interaction.

We begin our analysis with the case of spinless fermions and then
extend it to the case of spin-$1/2$ fermions.
We consider a system with a general Hamiltonian
\begin{eqnarray}
\!\!H\!=\!\!\sum_n\! U_n a_n^{\dagger}\! a_n \!+\!\!\!\sum_{m\neq n}\!\!T_{nm}a_n^{\dagger}\! a_m \!+\!\!\!\sum_{m\neq n}\!\!V_{nm}a_n^{\dagger}\!a_n a_m^{\dagger}\!a_m,
\label{Hfull}
\end{eqnarray}
where $a_n$ and $a_n^{\dagger}$ are fermionic creation and annihilation operators corresponding 
to different orbitals or fermionic sites in lattice models. This Hamiltonian generalizes many models: the Hubbard model,
finite-difference discretization in quantum chemistry, double-basis
coding, etc. 

The method of Ref. \cite{FSG} uses the well known Jordan-Wigner
transformation \cite{JORDAN-WIGNER_TRANSFORMATION} as well as 
the Trotter-Suzuki decomposition \cite{Trotter,Suzuki} of the evolution operator. A
key idea is that qubit numbering in Jordan-Wigner transformation can
be changed on-fly and this operation
can be combined with the action of the evolution operator. The resulting
operation is represented in terms of the so-called two-qubit fermionic
simulation gate \cite{FSG} (FSG) parameterized by the Hamiltonian constants. 
In Appendix A we show how FSG can be represented in terms of the native gates.
We also discuss the general structure of the quantum circuit for both spinless and spin-$1/2$ fermions.

\section{Fermionic SWAP network for different connectivity topologies}

A quantum circuit presented in terms of standard quantum gates (for example, CNOTs) can be recast through the native operators (Appendix A). However, these operators must be executed only between qubits affected by CNOTs. Thus, an important task is to disable the interaction between certain qubits, which can be achieved using the so-called refocusing operations \cite{refocusing1,refocusing2,refocusing3}. The need for a refocusing operation occurs each time a set of two-qubit gates must be applied simultaneously to different pairs of qubits (for example, CNOTs). 

\begin{figure}[ht]
\includegraphics[width=0.6\linewidth]{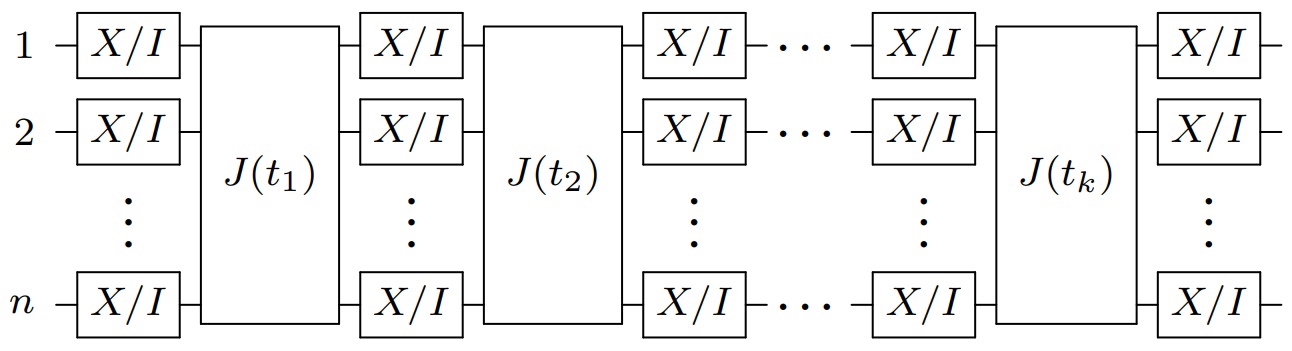}
\caption{General layout of the refocusing operation. It contains layers of single-qubit operations $X$ or $I$ interleaved with the action of multi-qubit entangler $J$. Each qubit is characterized by its sequence of operations $X$ and $I$.}
\label{Balance}
\end{figure}

We now explain the general principle of engineering the refocusing operations when implementing a fermionic SWAP network. A refocusing operation is based on a partition of the total interaction time. The total time is determined by the desired operators we wish to realize. For example, for CNOT gates this dimensionless time is $\pi/4$ (Appendix A). For Cphase gates applied simultaneously, such a duration is determined by the maximum single-qubit rotation, which characterizes the given set of Cphase gates. The refocusing operation also involves the action of the Pauli operators $X$ between some of the idling times. Any refocusing operation can be understood in terms of the action or non-action of $X$ between the times of the $t_i$ interaction, as shown in Fig. \ref{Balance}. In our work, $\sum_{i=0}^kt_i=T$ and $t_i=T/k$, where $T$ is the total interaction time. The rule of application of $X$ and $I$ gates will be explained in the next Subsection.

The unitary transformation with the operator $X$ in Fig. \ref{Balance} corresponds to the reversed time for a given qubit: 

\begin{eqnarray}
X_p \exp(-iZ_pZ_qt)X_p=\exp(iZ_pZ_qt) .
\label{invertevol}
\end{eqnarray}

Thus, using $X$ gates during the refocusing operation allows the implementation of a desired multi-qubit gate. 
However, if there are several such transformations in a row, there are double applications of $X$ gates, which can be replaced by the action of the operator $I=XX$. For the sake of convenience, we will say that within a given refocusing operation, the qubit is in an even-parity (odd-parity) state if the operator $X$ acted on the qubit an even (odd) number of times. Thus, the description of odd- and even-parity states is equivalent to the description with the time-reversed.

\subsection{All-to-all connectivity}

In this Subsection, we consider such quantum networks for which any qubit interacts with all other qubits. Suppose that our goal is the implementation of effective interaction between only qubits $i$ and $j$ over time $T$. The minimum time of the refocusing operation coincides with $T$. In order to achieve the goal, both qubits $i$ and $j$ must be either in the even-parity state, or in the odd-parity state at all steps of the refocusing operation. For the sake of convenience, we will refer to such pairs of qubits as interaction pairs. 
For example, the number of such qubits pairs for the fermionic SWAP network is $\frac{n}{2}$ for an odd layer of FSG operators and $\frac{n}{2}-1$ for an even layer, where $n$ is the number of qubits (see, e.g., Fig. \ref{FSGlay} of Appendix A). It is necessary not only to implement the desired operation within each interaction pair. It is also necessary to disable interaction between qubits from different pairs, as well as between all qubits and those qubits that are not included in any of the interaction pairs. We will refer to such qubits as single qubits. An interaction pair is in an odd-parity (even-parity) state if each qubit of the pair is in an odd-parity (even-parity) state. 

In our approach, the number of time intervals $t_i$ in the refocusing operation is always $m$, and all these intervals are equal, where $m$ is a natural number satisfying a condition
\begin{eqnarray}
k\leq m < 2k,
\label{condition}
\end{eqnarray}
where $k$ is the total number of interaction pairs and single qubits, determined by the type of quantum operation we wish to accomplish. Let us stress that Eq. (\ref{condition}) is correct only for the case of all-to-all connectivity. 

Appendix B presents the mathematical formalism behind the concept of the refocusing operation including the Hadamard operator $H(m)$, which plays a crucial role \cite{refocusing1}. It is defined in a matrix form as
\begin{eqnarray}
H(m)H(m)^T=mI,
\end{eqnarray}
where $I$ is identity matrix and $H (m)$ is the square matrix $m\times m$, with elements $\pm 1$. For instance, $H(2)=\left(\begin{array}{cc}
1 & 1 \\
1 & -1\\
\end{array}
\right).$

An important property of the operator $H(m)$ is the fact that any two columns of this operator in a matrix form, if they are considered as two vectors, are orthogonal to each other, and absolute values of any element are the same. This property is used for the construction of the refocusing operation \cite{refocusing1,refocusing2,refocusing3}: the operator $H(m)$ defines the sequences of states (even- or odd-parities) for every single qubit or interaction pair, which they must follow. Every single qubit or interaction pair can be associated with an arbitrary column up to the mutual permutations. The element $i$ of column $j$ determines the state (even or odd parity) of an interaction pair or single qubit with index $j$ during $t_i$. If this element is equal to $1$, then the qubit pair or single qubit must be in the even-parity state, and if $-1$, then it must be in the odd-parity state. As a result, each matrix column defines a set of states. If the qubit follows the sequence of these states during the refocusing operation, it does not effectively experience interaction with the qubits of other groups. If our goal is to implement CNOT operations simultaneously between certain qubits, then, in the beginning, we must find an appropriate $m$ for a given connectivity topology. In the case of all-to-all connectivity, $m$ is defined according to Eq. (\ref{condition}). The algorithm for determining $m$ in the case of a general connectivity topology is presented in Appendix C. 

Let us estimate a circuit depth, i.e., the number of multi-qubit gate applications per single Trotter step per qubit. We start with the special case, when qubit number $n$ can be represented as $\frac{n}{2}+2=2^\mathcal{N}$, where $\mathcal{N}$ is a natural number. From the viewpoint of resources, this is an ideal case for a digital-analog quantum device with full connectivity applied for the simulation of the quantum system described by the Hamiltonian (\ref{Hfull}). It is known that $H(m)$ exists for $m=\frac{n}{2}+2$. The number of rows of the square matrix $H(m)$ corresponds to the number of $t_i$ for an individual refocusing operation. In this case, the number of applications of interaction gates $J$ is $\frac{3n^2}{2}+6n$ per Trotter step. The number of CNOTs needed to implement the same Trotter step on a digital quantum computer is $\frac{3n^2-3n}{2}$ (the number of FSGs is $\frac{n^2-n}{2}$, while each FSG is represented in terms of three CNOTs). In both cases, the number of multi-qubit gates is $\mathcal{O}(n^2)$, which is preserved for any $n$. However, the circuit depth for a digital-analog device is $\mathcal{O}(n^2)$, and for a digital one, it is $\mathcal{O}(n)$. This difference is significant, so the use of a digital-analog quantum device is not optimal for implementing the fermionic SWAP network in the case of the full connectivity of qubits. Next, we will consider the connectivity, for which the circuit depth will be the same for digital and digital-analog quantum devices.

\subsection{Two-dimensional topology}

\begin{figure}[ht]
\center{\includegraphics[width=0.3\linewidth]{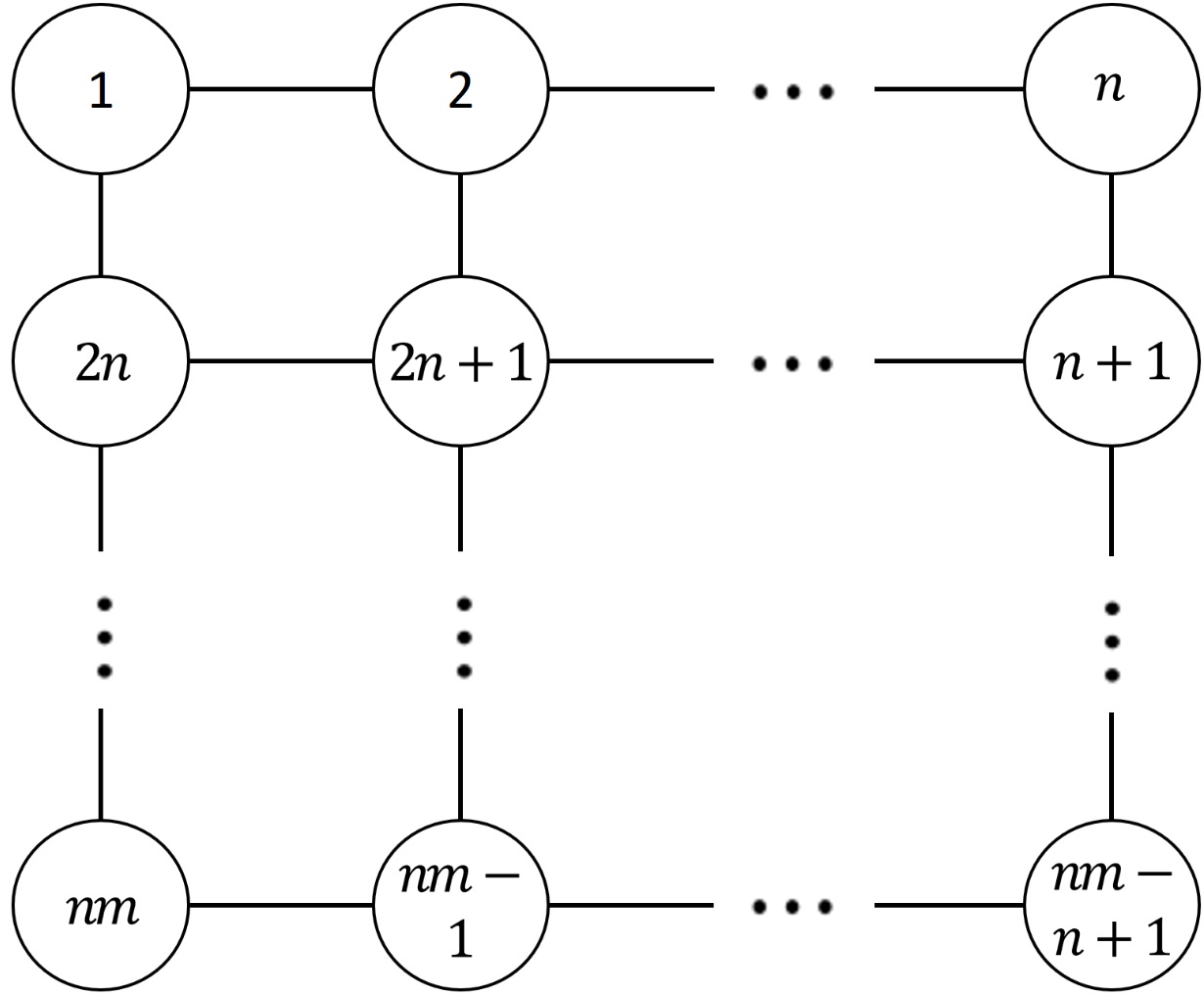}}
\caption{ Two-dimensional topology of qubit connectivity: square lattice with nearest-neighbor interaction. Lines connecting qubits show the action of the operator $J_{pq}$ between them.}
\label{Twodimensional}
\end{figure}

In this Subsection, we discuss two-dimensional connectivity topologies. An example, represented by the square lattice with nearest-neighbor interaction, is shown schematically in Fig. \ref{Twodimensional}.

\begin{figure}[ht]
\hfill
\center{\includegraphics[width=.7\linewidth]{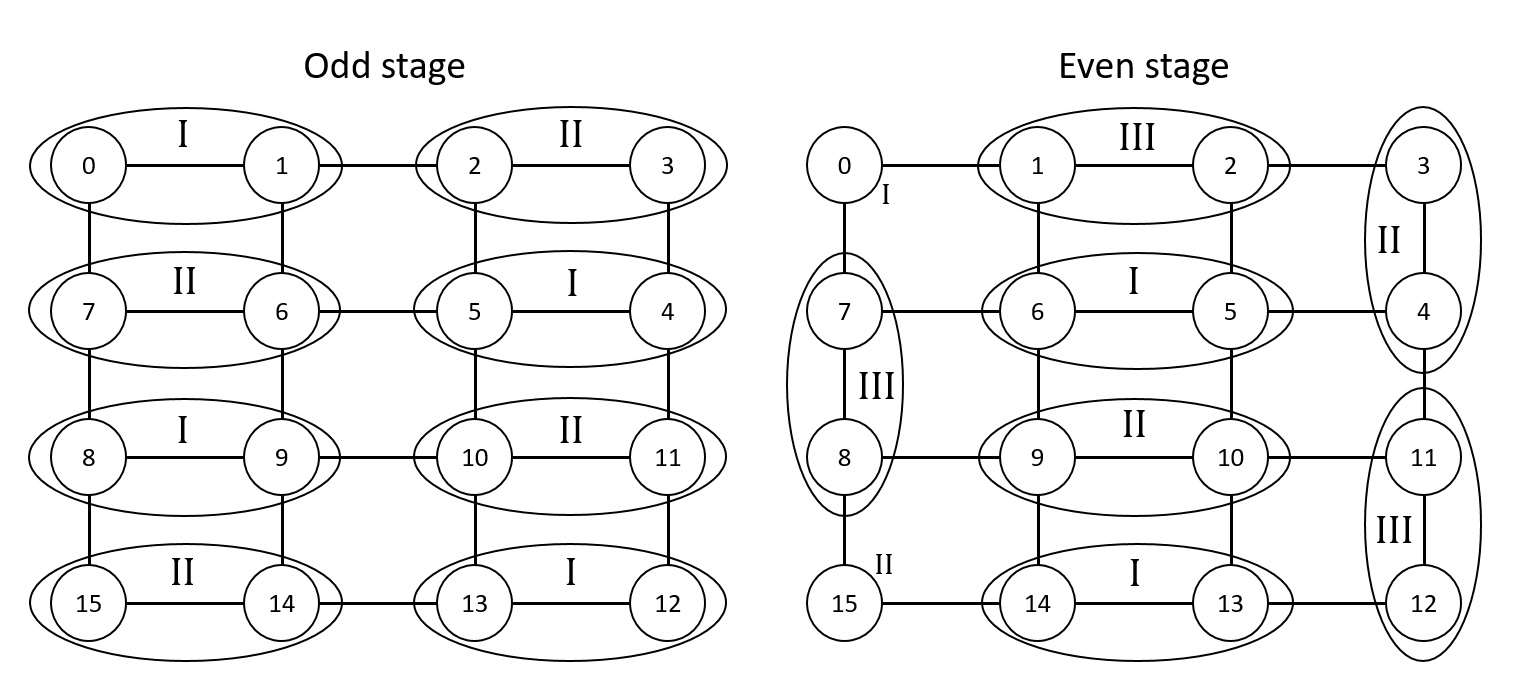}}
\caption{ General layout for the implementation of a single Trotter step (see in the text). }
\label{Trottertwodimensiona}
\end{figure}

Consider, for example, the qubits $1$ and $n+2$ in Fig. \ref{Twodimensional}. Since the native interaction does not act between them, these two qubits will not interact during the refocusing operation, no matter what sequences they follow. The fact that the qubit interaction is limited reduces the number of iterations $t_i$ required for the refocusing operation. We will demonstrate this fact using the example shown in Fig. \ref{Trottertwodimensiona}. The qubits included in the interaction pair are circled. The odd stage of the procedure is used for the odd layers of FSG operators, and even stage for the even layers. These stages correspond to the single Trotter step for simulating the system described by the Hamiltonian (\ref{Hfull}). Note that the qubit numbering, as indicated in Fig. \ref{Trottertwodimensiona}, and the fermionic site numbering generally do not match. For an odd stage, the refocusing operation can be performed using only a two-qubit behavior sequence defined by the matrix $H(2)$ that dramatically simplifies the problem compared to the case of all-to-all connectivity. In order to implement the refocusing operation at odd stages, interaction pairs must be positioned in a staggered pattern order, as depicted in Fig. \ref{Trottertwodimensiona}. In this case, qubit pairs indicated by "I" follow the first sequence, while qubit pairs indicated by "II" follow the second sequence. Since only two-qubit behavior sequences are used, the number of time intervals $t_i$ is two, where $t_i=\frac{T}{2}$. 

In the case of even stages, also shown in Fig. \ref{Trottertwodimensiona}, the sequences are determined by $H(4)$, which encodes four possible sequences of states. Two sequences, in this case, are not sufficient. Indeed, let us consider the interaction pairs formed by qubits $1-6$ at the even stage in Fig. \ref{Trottertwodimensiona}. Two sequences are not sufficient since there are triples of interaction pairs, where all three interact with each other. According to Appendix C, four sequences of states are enough for this connectivity topology.  Moreover, three sequences can be used in the even stage, and the association between the qubits and the three sequences is shown in Fig. \ref{Trottertwodimensiona} by "I", "II", and "III". However, $H(3)$ does not exist, which leads us to use the three sequence of states generated by $H(4)$.

The circuit depth for the implementation of a single Trotter step grows as $12n$, i.e., it scales as $\mathcal{O}(n)$. This estimate coincides with a similar result for a digital quantum device. In the latter case, the total number of CNOT gates is $\mathcal{O}(n^2)$. We conclude that a digital-analog quantum device with a two-dimensional topology is prospective for the quantum simulation of fermionic models.

\subsection{One-dimensional topology}

\begin{figure}[ht]

\center{\includegraphics[width=.5\linewidth]{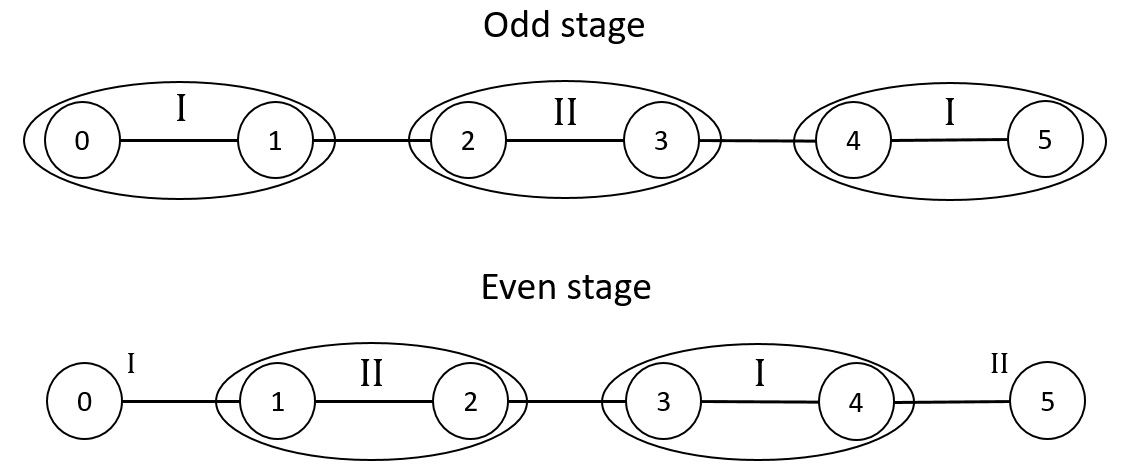}}
\caption{Scheme of linear connectivity of qubits. The interaction pairs at even and odd stages of a single Trotter step are shown. }
\label{Trotteronedimensional}
\end{figure}

The one-dimensional connectivity topology is the simplest case of qubit network connectivity. It turns out that this is an ideal case of using the fermionic SWAP network. For the one-dimensional topology, two sequences of states are sufficient so that the appropriate sequences of states are determined by the matrix $H(2)$. The association between qubits of the chip and interaction pairs at even and odd stages of the single Trotter step is shown in Fig. \ref{Trotteronedimensional}. This association takes an elementary form. We remind that a single Trotter step consists of $n/2$ odd stages and $n/2$ even stages. The circuit depth grows as $6n$, and this value is twice smaller than that for the case of the two-dimensional square qubit array due to the number of $t_i$ used in the refocusing operation. In the case of one- and two-dimensional qubit networks, the circuit depth required to implement a single layer of FSGs is $\mathcal{O}(1)$, since the number of sequences of states required to implement one layer of FSGs is constant for any qubit number $n$, according to Appendix C.

Note that using quantum chips with one-dimensional connectivity usually dramatically increases the two-qubit gate number due to the necessity to utilize multiple SWAP operations. Surprisingly, quantum simulations of fermionic models in any dimension does not belong to this case due to the one-dimensional character of Jordan-Wigner transformation and despite the non-locality of the fermionic statistics. However, it turns out that for spin-$1/2$ fermions, the optimal connectivity topology of the quantum network is a ladder, as shown in Appendix D.

We also wish to stress that the general structure of the fermionic SWAP network is appropriate for the digital-analog strategy 
since all qubits of the device are used uniformly in time. The quantum circuit is 
represented by periodically accomplished layers of FSGs, which are executed each time for nearly half of qubits. 
Thus, multiple two-qubit entangling gates must be implemented simultaneously. This structure of the quantum circuit maximizes the usage of global entangling gates.

\section{Systems with a spread in interaction constants}

In this Section, we consider digital-analog quantum devices, which have some spread in interaction constants of qubits, and extend the refocusing technique to this important case that can be relevant for solid-state qubits, such as superconducting macroscopic artificial atoms, for which a precise control of chip parameters is difficult. The operator $e^{- i\alpha_{pq}Z_pZ_qt} $ acts on pairs of qubits with different coefficients $\alpha_{pq}$ for each qubit pair ($p$, $q$), as imposed by the topology. We assume that all these interaction constant are known.

We introduce a notion of a global refocusing operation, which represents a sequence of individual refocusing operations, described in the preceding Section. 
The result of such an operation is assumed to be the action of the operator $e^{- i\alpha_{pq}Z_pZ_qT_{pq}}$ only between pre-selected qubits, while for any two interaction pairs $ \alpha_{pq}T_{pq}=\alpha_{km}T_{km}$. We are interested in implementation of such an operator for each interaction pair, since when executing CNOT gates simultaneously between certain qubits, our goal is the same two-qubit operator $e^{-iZ_pZ_qT}$ with a fixed time $T$ for each pair $(p,q)$.

Without loss of generality, we assume that $\alpha_{01}$ is minimum among coupling constants, then we reassign the time $t^\prime=t\alpha_{01}$. The operator $e^{-i\beta_{pq}Z_pZ_qt^\prime}=e^{- i\alpha_{pq}Z_pZ_qt}$ acts on pairs of qubits according to the connectivity, where $\beta_{pq}=\alpha_{pq}/ \alpha_{01} \geq 1 $ so that $\beta_{01}=1$. From the condition $ \alpha_{pq}T_{pq}=\alpha_{km}T_{km}$, it follows that $T_{01}\geq T_{pq}$ for any ($p$, $q$). We construct a global refocusing operation with the duration $T=T_{01}$. Then the qubits $0$ and $1$ must interact during the time $T$. For all other interaction pairs, the same result is achieved in shorter time.  Thus, for each interaction pair, there is an excess time interval $\tau_{pq}=T-T /  \beta_{pq}$. This means that in the global refocusing operation, there is a time interval $\tau_{pq}$ for which the interaction between the qubits $p$ and $q$ must be disabled. This can be achieved by constructing a special refocusing operation, which is a part of the global refocusing operation.

Note that, within each individual refocusing operation, qubits characterized by different sequences of states do not interact with each other, since all time intervals $t_i$ are equal to each other. Hereafter, for the sake of simplicity, we use indices for numbering interaction pairs and single qubits (not individual qubits as before). Index $i=1$ will be used for the interaction pair with minimum interaction constant.

\subsection{General strategy}

We start our consideration with a relatively simple case, when the difference between $ \alpha_{i}$ is small enough, so that $\sum_{i}\tau_{i}< T$. 
Then the total interaction time can be divided into intervals $T_{i}=\tau_i$, $i\neq 1$, and $T_1=T-\sum_{k=2}^{n_p}\tau_k$, where $n_p$ is the number of interaction pairs. During each interval $T_i$, different interaction pairs must correspond to different sequences, then after this time, the interaction between qubits of different sequences is disabled, so they remain effectively non-interacting. For each interaction pair $i$ (except $i=1$), there must be a time interval $T_{i}$ when qubits from this interaction pair must remain non-interacting in order to compensate a difference in interaction constants. Hence, qubits of pair $i$ must follow different sequences of states during $T_i$ (except $i=1$) and the algorithm presented in Appendix B is again applicable for each $T_i$. 

The condition $\sum_{i}\tau_{i} < T$ is quite difficult to be fulfilled for a large number of qubits in the system. However, this condition is used for demonstration purposes only. Several interaction pairs can be destroyed during some of the time intervals $T_i$. Consider, for example, the case of an overlap of two times $\tau_i$ and $\tau_j$. Suppose that $\tau_i>\tau_j$, then in the global refocusing operation $T_j=\tau_j$, and $T_i=\tau_i-\tau_j$. This means that during $T_j$, the refocusing operation is constructed so that both interaction pairs $i$ and $j$ are destroyed (all four qubits become single), and during $T_i$, only the pair $i$ is destroyed (pair of qubits becomes single). Thus, the time intervals $\tau_i$ can overlap, which means that the quantum chip can be characterized by any coefficients $\alpha_{pq}$.

\subsection{Different topologies}

Let us consider a digital-analog quantum device with qubits positioned in the chain (linear connectivity). 
As an example, we analyze the system of six qubits that can be readily extended to the general case. 
We show how to construct an odd stage, shown in Fig. \ref{Trotteronedimensional}. 
The even stage is constructed similarly. 
The coupling between qubits 0 and 1 is again assumed to be minimum among all pairs of qubits. 
As it was shown earlier, two sequences are sufficient to implement the refocusing operation. There are three interaction pairs, and to achieve the result, we divide the time $T$ into three intervals as $T=T_1+T_2+T_3$. During each interval $T_1$, $T_2$, $T_3$ one of the interaction pairs is destroyed, except for $T_1$. 

Table 1 shows a sequence selection for six qubits. This table demonstrates what sequence each qubit must follow, according to the general algorithm presented in the preceding Subsection. In total, two sequences are used, and therefore, for each $T_i$, the refocusing operation contains only two intervals $t_i=T_i/2$. The result of such a global refocusing operation allows the implementation of a set of CNOTs between interaction pairs shown in Fig. \ref{Trotteronedimensional} for six qubits. Our general algorithm allows extending this approach to any qubit number for the one-dimensional topology.

\begin{figure}[ht]
\center{
\begin{tabular}{|l|l|l|l|l|l|l|}
\hline
Q   & 0 & 1 & 2 & 3 & 4 & 5 \\ \hline
$T_1$ & 1 & 1 & 2 & 2 & 1 & 1 \\ \hline
$T_2$ & 1 & 1 & 2 & 1 & 2 & 2 \\ \hline
$T_3$ & 1 & 1 & 2 & 2 & 1 & 2 \\ \hline
\end{tabular}
}
\caption*{\textbf{Table 1.} A sequence of states to achieve a fixed action of the operator $e^{-iZ_pZ_qT}$ between all interaction pairs for the case of linear connectivity of qubits with the spread in interaction constants. The number in the table shows which sequence, given by the matrix $H(m)$, the qubit is associated.
}
\end{figure}

\begin{figure}[ht]
\center{\includegraphics[width=0.44\linewidth]{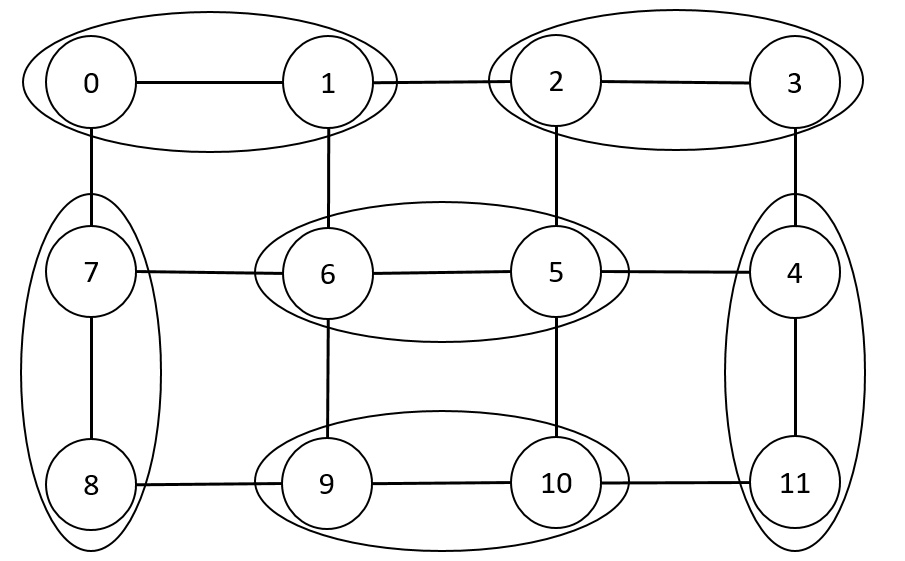}}
\caption{Schematic layout of the two-dimensional square qubit network.}
\label{disorderedtwodimensional}
\end{figure}

We now consider a digital-analog quantum device with two-dimensional connectivity of qubits, as shown in Fig. \ref{disorderedtwodimensional}. In this case, each qubit can interact with no more than four neighboring qubits. Therefore, four sequences defined by the matrix $H(4)$ are sufficient for refocusing. They are shown in Table 2. This example is interesting because the qubits 5 and 6, which form the interaction pair, interact with five other interaction pairs at once. However, according to our general algorithm, four sequences are sufficient to implement the desired operator between the selected qubits circled in Fig. \ref{disorderedtwodimensional}. Note that the selection of sequences is not unique up to permutations in the same way as the system with identical interaction constants.

\begin{figure}[ht]
\center{
\begin{tabular}{|l|l|l|l|l|l|l|l|l|l|l|l|l|}
\hline
Q   & 0 & 1 & 2 & 3 & 4 & 5 & 6 & 7 & 8 & 9 & 10 & 11 \\ \hline
$T_1$ & 1   & 1   & 2   & 2   & 4   & 3   & 3   & 2   & 2   & 1   & 1    & 4    \\ \hline
$T_2$ & 1   & 1   & 2   & 1   & 4   & 3   & 3   & 2   & 2   & 1   & 1    & 4    \\ \hline
$T_3$ & 1   & 1   & 2   & 2   & 4   & 3   & 4   & 2   & 2   & 1   & 1    & 4    \\ \hline
$T_4$ & 1   & 1   & 2   & 2   & 4   & 3   & 3   & 2   & 4   & 1   & 1    & 4    \\ \hline
$T_5$ & 1   & 1   & 2   & 2   & 4   & 3   & 3   & 2   & 2   & 1   & 2    & 4    \\ \hline
$T_6$ & 1   & 1   & 2   & 2   & 4   & 3   & 3   & 2   & 2   & 1   & 1    & 2    \\ \hline
\end{tabular}
}
\caption*{\textbf{Table 2.} A sequence of states needed to achieve a fixed action of the operator $e^{-iZ_pZ_qT}$ between all interaction pairs for the case of two-dimensional connectivity of qubits with a spread in the interaction constants.
}
\end{figure}

\subsection{Remarks on scaling}

When the spread of parameters is considered, the circuit depth for a single Trotter step increases by a factor of $\mathcal{O}(n)$ and becomes equal to $\mathcal{O}(n^2)$. This is due to the fact that the refocusing operations have to be replaced by global refocusing operations. Each global operation consists of $\mathcal{O}(n)$ individual refocusing operations. The number of multi-qubit operations also grows as $\mathcal{O}(n^2)$. 

The general idea of engineering a global refocusing operation is also valid for the all-to-all connectivity topology of qubits. However, as noted earlier, this connectivity is not effective for the implementation of the fermionic SWAP network since each refocusing operation contains $\mathcal{O}(n)$ time intervals $t_i$. Thus, the number of multi-qubit gates for the single Trotter step grows as $\mathcal{O}(n^3)$.

\section{Error analysis}

In this Section, we consider the impact of two types of errors, which are important for digital-analog quantum computing. One of these errors is due to the the inaccurate determination of $\alpha_{pq}$, which differs from the actual value, for the entangling operator $e^{- i\alpha_{pq}Z_pZ_qt} $ between qubits $p$ and $q$. Another error we consider is the depolarizing noise, which is here described in terms of the action of Kraus operators after each application of $J(t)$ gate.

To describe the effect of errors, we use the concept of fidelity of quantum processes defined as
\begin{eqnarray}
F_{proc}(A,B)=\frac{\vert Col(A)^\dagger Col(B)\vert^2}{Col(A)^\dagger Col(A)Col(B)^\dagger Col(B)},
\label{Fidelitydefinition}
\end{eqnarray}
where $Col$ is the operation of converting a matrix into a column:
\begin{eqnarray}
Col\left(\begin{array}{cc}
a & b \\
c & d\\
\end{array}
\right)=\left(\begin{array}{c}
a\\
c\\
b\\
d\\
\end{array}
\right).
\label{Col_definition}
\end{eqnarray}
In order to estimate the difference between the results of the ideal and not ideal simulations, we use fidelity of quantum states defined as
\begin{eqnarray}
F(\sigma,\rho)=\left(tr(\sqrt{\sqrt{\rho}\sigma\sqrt{\rho}})\right)^2,
\label{fidelity_of_states}
\end{eqnarray}
where $\sigma$ and $\rho$ are the density matrices of the quantum states.

\subsection{Error of inaccurate determination of the coupling constant}

In this Subsection, we analyze an impact of the error of inaccurately defining the constant $\alpha_{p q}$ for the entanglement operator $e^{-i\alpha_{pq}Z_pZ_qt}$ between qubits $p$ and $q$. Suppose we were assuming that the operator $e^{-i\alpha_{pq} Z_pZ_qt}$ acts between them, but in reality the interaction constant was $\alpha_{pq}+\alpha_{pq}^\prime$. Thus, an error occurs associated with nonzero $\alpha^\prime_{pq}$.

\begin{figure}[ht]
\includegraphics[width=0.5\linewidth]{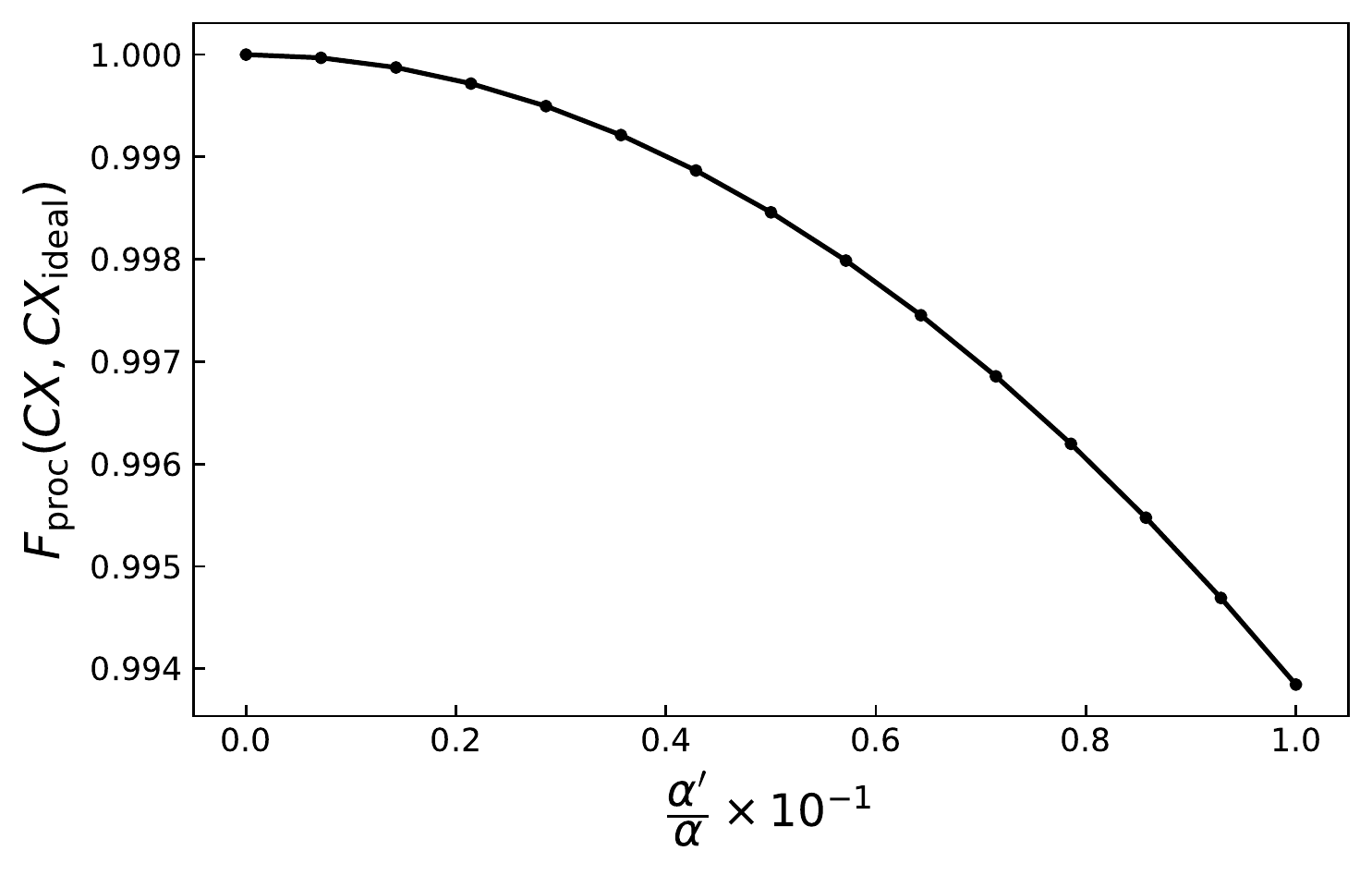}
\caption{The dependence of $F_{\text{proc}}(CX,CX_{\text{ideal}})$ on $\alpha^\prime/\alpha$.}
\label{Fcnot}
\end{figure}

We start our consideration with an estimate of $F_{proc}$ for the CNOT execution for two isolated qubits for which the actual interaction constant is $\alpha+\alpha^\prime$ instead of $\alpha$. The fidelity of quantum process value introduced in Eq. (\ref{Fidelitydefinition})  can be expressed explicitly as
\begin{eqnarray}
F_{\text{proc}}(CX,CX_{\text{ideal}})=\cos^2\left(\frac{\pi}{4}\frac{\alpha^\prime}{\alpha}\right).
\label{F_cnot}
\end{eqnarray}
The infidelity $1-F_{\text{proc}}(CX,CX_{\text{ideal}})$ scales as $\frac{1}{2}(\frac{\pi}{4}\frac{\alpha'}{\alpha})^2$ at $\alpha' \ll \alpha$. Thus, only a weak quadratic growth of infidelity is expected, as $\alpha'$ grows. Figure \ref{Fcnot} presents the numerically obtained dependence (dots) of $F_{\text{proc}}(CX,CX_{\text{ideal}})$ on $\alpha^\prime/\alpha$, which shows a full agreement with the explicit result (solid line). A weak dependence of infidelity on $\alpha'$ is a positive fact in the view of perspectives of the digital-analog approach.

\begin{figure}[ht]
\includegraphics[width=0.5\linewidth]{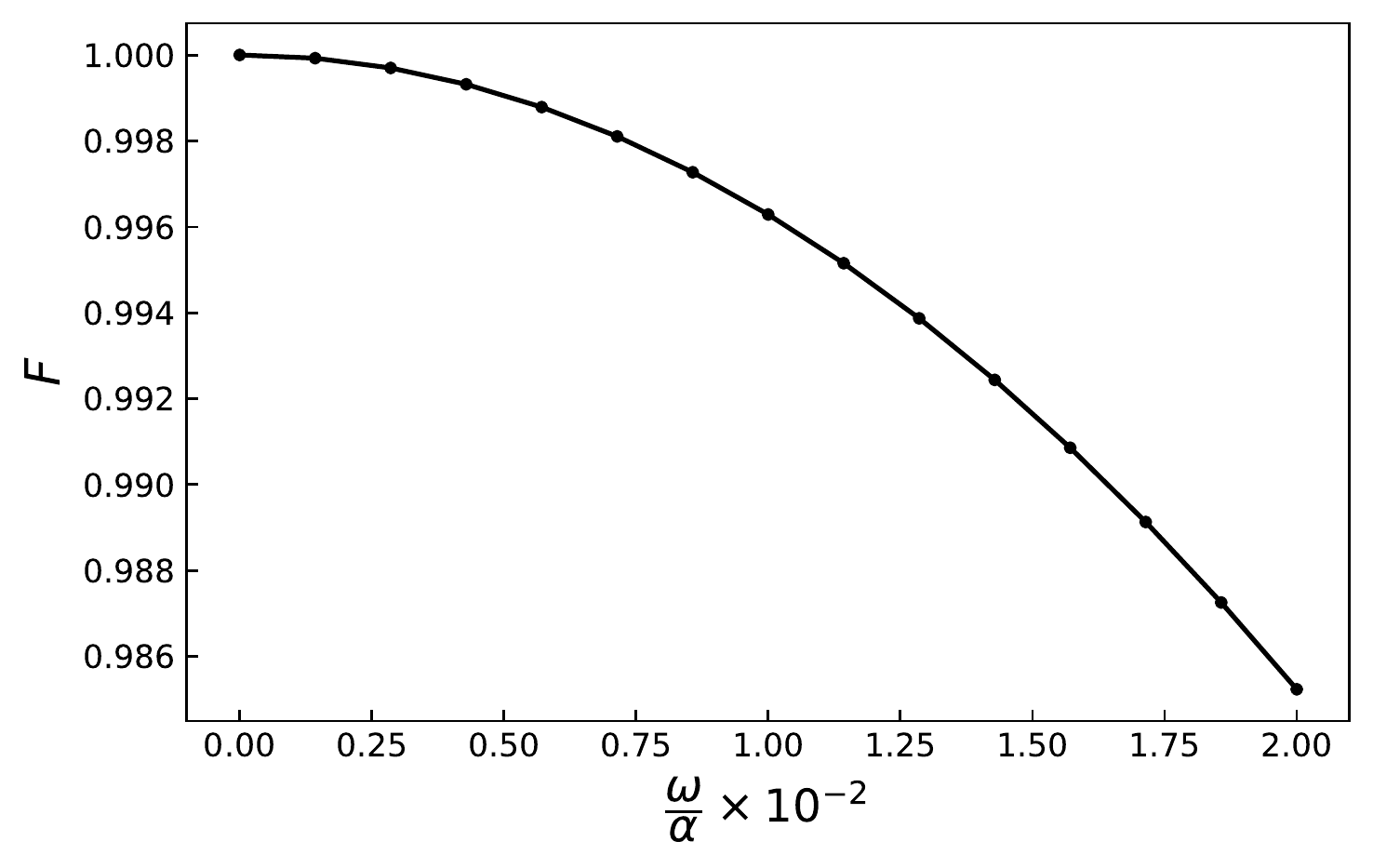}
\caption{The dependence of $F$ on $\omega/\alpha$ for a single Trotter step and 12 qubits.}
\label{F_trotter}
\end{figure}

Our next step is a consideration of the impact of an error in the definition of coupling constants for the single Trotter step for the Hamiltonian (\ref{Hfull}). We consider the case when a deviation of the coupling constant is random for any pair of neighboring qubits. The most optimal linear connectivity topology of the qubit network is assumed. 
We assume they are randomly selected from a uniform distribution from the segment $[-\omega,\omega]$. 
Figure \ref{F_trotter} shows the dependence of $F$ of an ideal Trotter step simulation and a simulation with errors on $\omega/ \alpha$ 
for the chain of twelve qubits. This dependence was found numerically for different values of $\omega/ \alpha$, and each value corresponds to a dot in Fig. \ref{F_trotter}. According to the Haar measure, a random state is selected as the initial state, and then averaging is performed for 100 initial states. 
The energy parameters $V_{nm}$ and $T_{nm}$ of the Hamiltonian (\ref{Hfull}) are randomly selected from a uniform distribution so that $V_{nm}\delta t$, $T_{nm}\delta t \in [-0.1,0.1]$ for any $n,m$ (all-to-all connectivity of the fermionic cluster). The results are presented for a particular realization of disorder. However, we found that they are essentially insensitive to the realizations. The coefficients $U_n$ are assumed to be zero since single-qubit gates simulate the corresponding evolution operator. We again see a weak quadratic growth of infidelity, now for the whole Trotter step, as a function of $\omega$.

Note that the time determination error is equivalent to the error in the determination of $\alpha$ since $\alpha$ and $t$ are symmetric for the entanglement operator $J(t)$.

\subsection{The effects of depolarizing noise, amplitude damping, and phase damping}

In this Subsection, we consider the impacts of three types of uncorrelated noise \cite{nielsen2002quantum} in the case of the quantum circuit corresponding to the single Trotter step. In our model, the density matrix of each qubit is transformed according to the noise model after each two-qubit operation, which is $J(t)$ and CNOT in the case of digital-analog and digital quantum computing, respectively. The transformations correspond to the action of the Kraus operators on the density matrix, given by
\begin{eqnarray}
\mathcal{E}^d_p(\rho)=(1-p)\rho+\frac{p}{3}X\rho X+\frac{p}{3}Y\rho Y+\frac{p}{3}Z\rho Z,
\label{depolar}
\end{eqnarray}

\begin{eqnarray}
\mathcal{E}^a_\gamma(\rho)=E_0^{a}\rho E_0^{a}+E_1^{a}\rho E_1^{a\dagger},\qquad E_0^a=\left(\begin{array}{cc}
1 & 0 \\
0 & \sqrt{1-\gamma}\\
\end{array}
\right), \qquad E_1^{a}=\left(\begin{array}{cc}
0 & \sqrt{\gamma} \\
0 & 0\\
\end{array}
\right),
\label{amplitude}
\end{eqnarray}

\begin{eqnarray}
\mathcal{E}^p_\lambda(\rho)=E_0^p\rho E_0^p+E_1^p\rho E_1^p,\qquad E_0^p=\left(\begin{array}{cc}
1 & 0 \\
0 & \sqrt{1-\lambda}\\
\end{array}
\right), \qquad E_1^p=\left(\begin{array}{cc}
0 & 0 \\
0 & \sqrt{\lambda}\\
\end{array}
\right),
\label{phase}
\end{eqnarray}
In the model corresponding to Eq. (\ref{depolar}) one of the Pauli errors occurs with probability $p$ giving rise to a completely mixed qubit state. 
The model (\ref{amplitude}) corresponds to the effect of energy dissipation, while the relaxation time is characterized by the parameter $\gamma$. 
The noise that describes the loss of information without energy loss (phase damping) is represented by the model (\ref{phase}).

\begin{figure}[h!]
\includegraphics[width=0.4\linewidth]{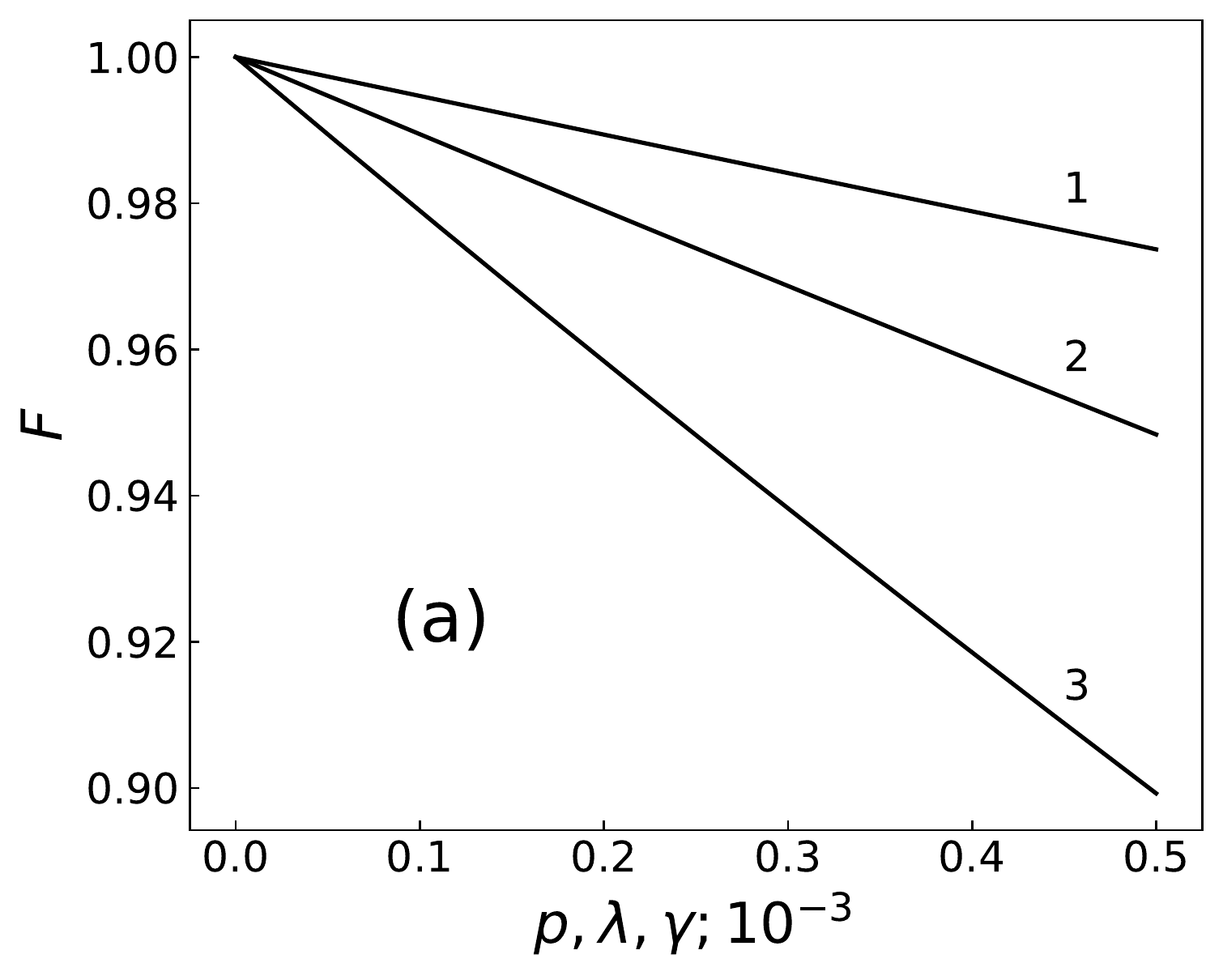}
\includegraphics[width=0.4\linewidth]{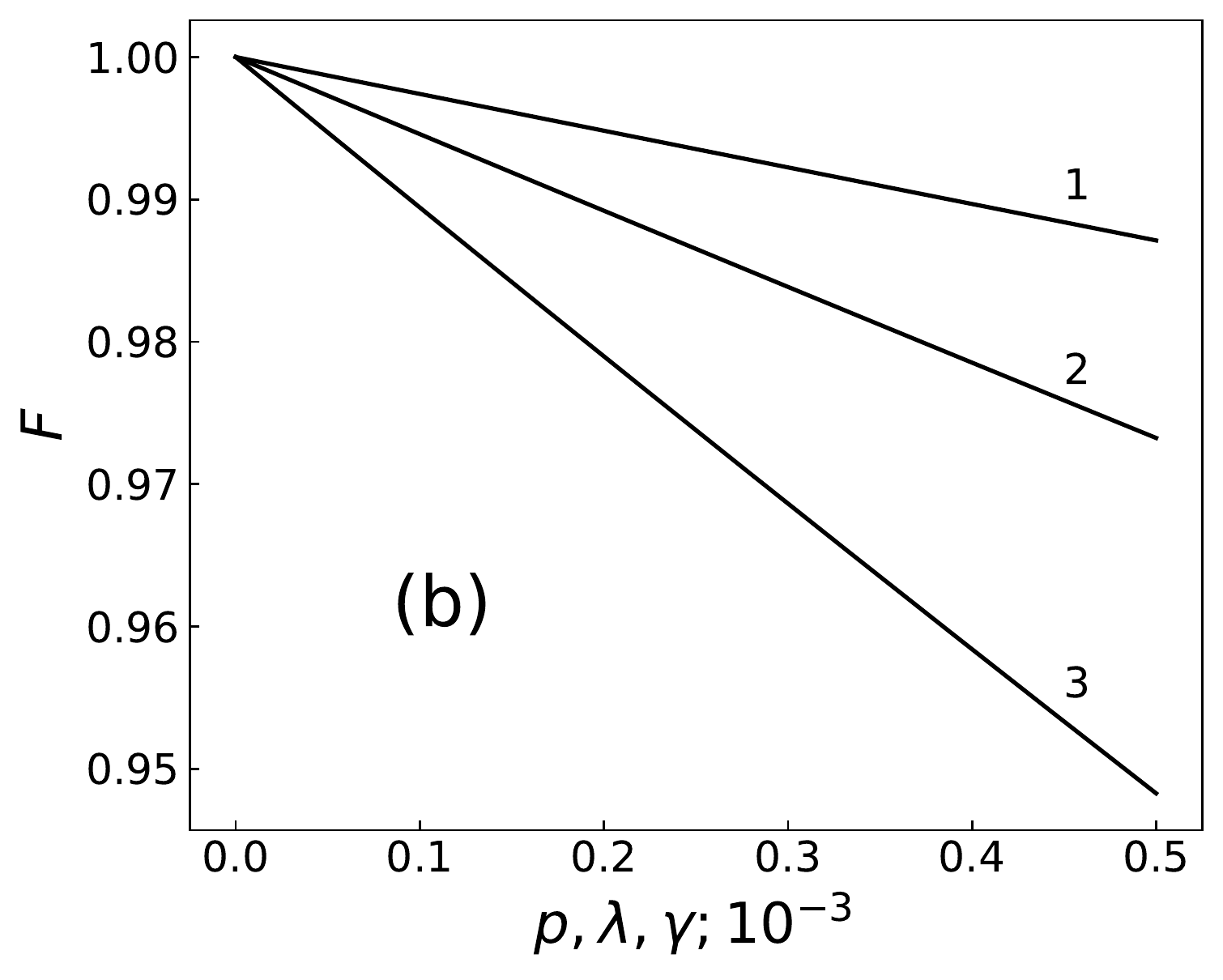}

\caption{The dependence of $F$ for a single Trotter step on error parameters for the quantum processor consisting of six qubits positioned in a chain. 
Two types of quantum devices are considered: digital-analog quantum computer (a) and digital quantum computer (b).
Three types of errors are analyzed: phase damping (curves 1), amplitude damping (curves 2), depolarizing noise (curves 3). }
\label{errors}
\end{figure}

We again consider the single Trotter step of evolution for the Hamiltonian (\ref{Hfull}) from the initial random state selected according to the Haar measure and then average the fidelity for 100 different initial states. The parameters of the fermionic model are the same as in the preceding Subsection. 
A linear connectivity topology of the quantum chip is assumed both in the digital-analog and digital cases. In the latter case, execution of CNOTs is assumed.
Figure \ref{errors} shows the dependence of the fidelity on the noise parameters for the chain of six qubits
computed numerically for the digital-analog (a) and the digital (b) models. 
In all cases considered, we see a pronounced linear decrease of fidelity as noise parameter grows. Thus, these noises are generally more dangerous for the simulation of fermionic systems than the error of inaccurate determination of the coupling constant within the digital-analog strategy. In addition, for each type of error, the value of infidelity in the case of digital-analog approach is twice as large as in the case of digital computation for the same values of noise parameters. This is directly related to the fact that two native entangling operations 
are used to implement a single refocusing operation according to linear connectivity topology.
We, however, stress that such a direct comparison of fidelities without specifying the implementation and architecture of a quantum computer is not reasonable, since actual values of noise parameters for both strategies can be very different even for the same physical realization of the quantum device and are generally determined by specific error budgets. 
However, in all cases, we observe a linear fidelity decrease, while the unique error for digital-analog strategy has only a weak quadratic effect on infidelity. 
At the same time, a key advantage of the digital-analog strategy is that it does not require two-qubit gates implementation that is associated with complicated engineering and wiring issues. 
This supports our vision that digital-analog quantum computing is attractive for quantum simulation of fermionic systems. 

We also would like to mention that Ref. \cite{Walter} presents simple estimates for the optimal coupling strength
for a digital-analog strategy based on always-on interaction, which appears as a result of a balance between the fidelities of single-qubit gates diminished by
always-on interaction and fidelities of entangling operations decreased due to the decoherence. Such a consideration can be helpful in construction of digital-analog devices
based on different physical realizations.

\section{Conclusions}

This article pointed out that digital-analog quantum devices are prospective for the quantum simulation of fermionic models in different dimensions. 
The approach we proposed is based on the application of fermionic SWAP network \cite{FSG} implemented through always-on interaction between qubits of the device. The fermionic simulation gate \cite{FSG} can be used to circumvent the non-locality problem of the Hamiltonian (\ref{HHubbard}) under the Jordan-Wigner transformation. This gate is represented through the native interaction of qubits and single-qubit gates. We suggested using a refocusing technique when implementing a fermionic SWAP network to compensate for undesirable interactions between qubits at each simulation step. 
We argued that the digital-analog strategy is appropriate for the fermionic simulation since all qubits of the chip can be used uniformly in time, enabling it to perform multiple entangling operations simultaneously. 

We concentrated mainly on the impact of the connectivity topology of the qubit network, its spread in interaction constants, and the effect of errors in entangling operations on the efficiency of quantum algorithm implementation. We found that an optimal topology for the digital-analog chip in the context of fermionic simulation is a one-dimensional chain for any dimensionality of the simulated model, in the case of spinless fermions, and a ladder, in the case of spin-1/2 fermions. In this case, the overhead associated with compensating undesired interactions between qubits is minimum. Such a simple connectivity topology makes it easier to probe our findings in experiments. 

For optimal topology, the simulation of the single Trotter step of the evolution requires $\mathcal{O}(n)$ applications of multi-qubit gates, where $n$ is the qubit number. The method of refocusing has been extended to the case when there is a spread of interaction constants. It is shown that the circuit depth increases by a factor of $\mathcal{O}(n)$. We also found that errors associated with the inaccuracy in determining interaction constants lead to the weak quadratic growth of infidelity as a function of the deviation of the actual coupling constants from the assumed values, which is a positive fact in the view of the perspectives of the digital-analog approach. The effects of the depolarizing noise, phase damping and amplitude damping are more harmful since they lead to the linear decrease of fidelity as a function of error parameter both for digital-analog and digital strategies.

\section*{Acknowledgements}
W. V. P. acknowledges a support from RFBR (project no. 19-02-00421).

\bibliography{references}

\appendix

\section{Fermionic SWAP network under always-on interaction}

\subsection{Spinless fermions}

Under Jordan-Wigner transformation for any numbering of spins, the first term of the
Hamiltonian (\ref{Hfull}) takes the form $\sum_p U_p\sigma_p^+\sigma_p$, where
$\sigma_p^+$ and $\sigma_p$ are the spin creation and annihilation
operators. The evolution operator
corresponding to this term can be modeled by the action of a
gate $R_z$  on each qubit and we do not
consider this contribution to the Hamiltonian.

\begin{figure}[ht]
\includegraphics[width=0.5\linewidth]{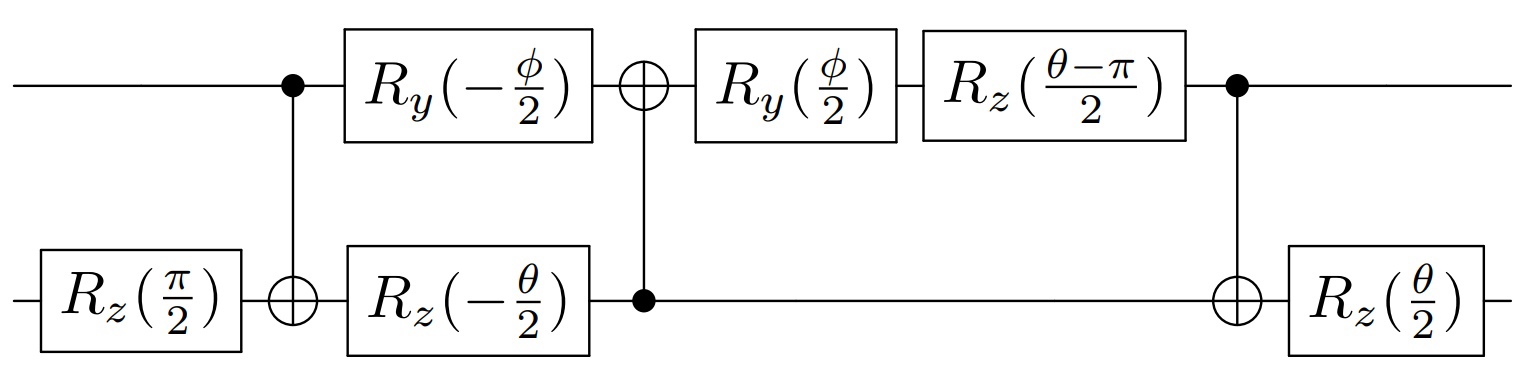}
\caption{Quantum circuit for FSG implementation in terms of single-qubit gates and CNOTs. } 
\label{ris:imageQC}
\end{figure}

Our goal thus is to simulate the evolution during the time interval
$\delta t$ under the Hamiltonian
\begin{eqnarray}\label{Htruncated}
H^\prime=\sum_{m\neq n}T_{nm}a_n^{\dagger} a_m + \sum_{m\neq n}V_{nm}a_n^{\dagger}a_n a_m^{\dagger}a_m.
\end{eqnarray}
As a result of the Jordan-Wigner transformation, the Hamiltonian (\ref{Htruncated})
takes an essentially non-local form. In order to circumvent this
problem, FSG gate can be utilized \cite{FSG,Obs_sep_dyn}, which is defined as 
\begin{eqnarray}
\begin{gathered}
\!\!\!\!F_{\delta t}^{nm}\!\!=\!e^{-iV_{pq}a^{\dagger}_na_na^{\dagger}_m a_m \delta
t}e^{-iT_{nm}(a^{\dagger}_n a_m+a^{\dagger}_ma_n) \delta
t}f_{swap}^{nm},
\end{gathered}
\end{eqnarray}
where $f_{swap}^{nm}=1+a_n^{\dagger}a_m+a_m^{\dagger}a_n-a_n^{\dagger}a_n - a_m^{\dagger}a_m$ is a
fermionic orbital change operation that preserves the correct
antisymmetrization. FSG is applied only for $m=n+1$, i.e., for
neighboring qubits. It can be represented using the matrix $F(\phi,\theta)$  defined as
\begin{eqnarray}
F(\phi,\theta)=\left(\begin{array}{cccc}
1 & 0 & 0 & 0\\
0 & -i\sin(\frac{\phi}{2}) & \cos(\frac{\phi}{2}) & 0\\
0 & \cos(\frac{\phi}{2}) & -i\sin(\frac{\phi}{2}) & 0\\
0 & 0 & 0 & e^{i \theta}\\
\end{array}
\right),
\end{eqnarray}
where $\phi/2=T_{nm} \delta t$ and $\theta = - V_{nm} \delta t+\pi$. 
Our strategy is to represent FSG in terms of
single-qubit gates and CNOTs and then express CNOT through the
native two-qubit gate. The implementation of FSG using CNOTs gates is shown in Fig. \ref{ris:imageQC}. 

\begin{figure}[ht]
\includegraphics[width=0.45\linewidth]{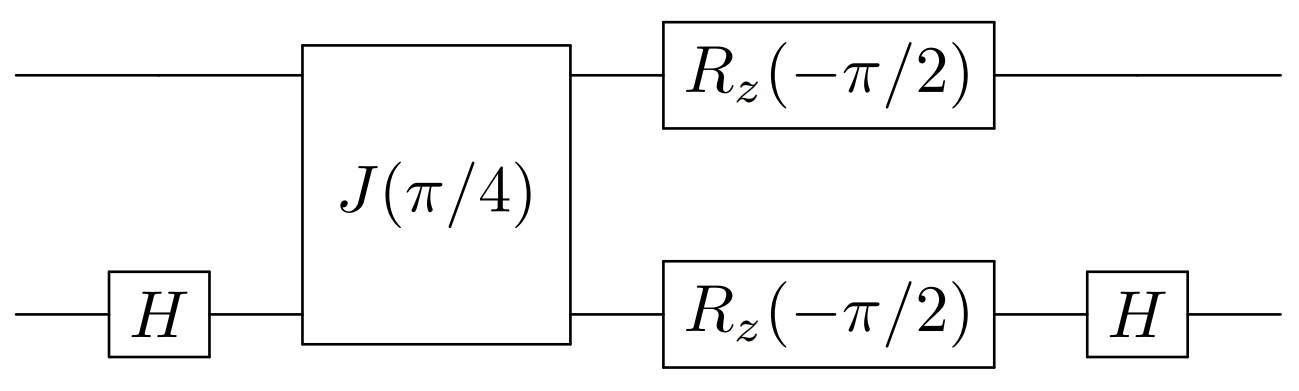}
\caption{ Implementation of  CNOT  gate on a
digital-analog quantum device. Control qubit corresponds to the upper 
line, while target qubit corresponds to the lower line.  } \label{CNOT}
\end{figure}

The implementation of CNOT gate on a digital-analog device is visualized in Fig. \ref{CNOT}: we consider an \emph{isolated} couple of qubits that experience native
interaction and express CNOT in terms of the quantum gate associated
with this interaction.  
A duration of always-on interaction is
$t_{CNOT}=\frac{\pi}{4}$. The refocusing technique used to effectively implement
the required isolation of a selected
set of couples is discussed in Sections III and IV of the main text.

\begin{figure}[ht]
\center{\includegraphics[width=0.4\linewidth]{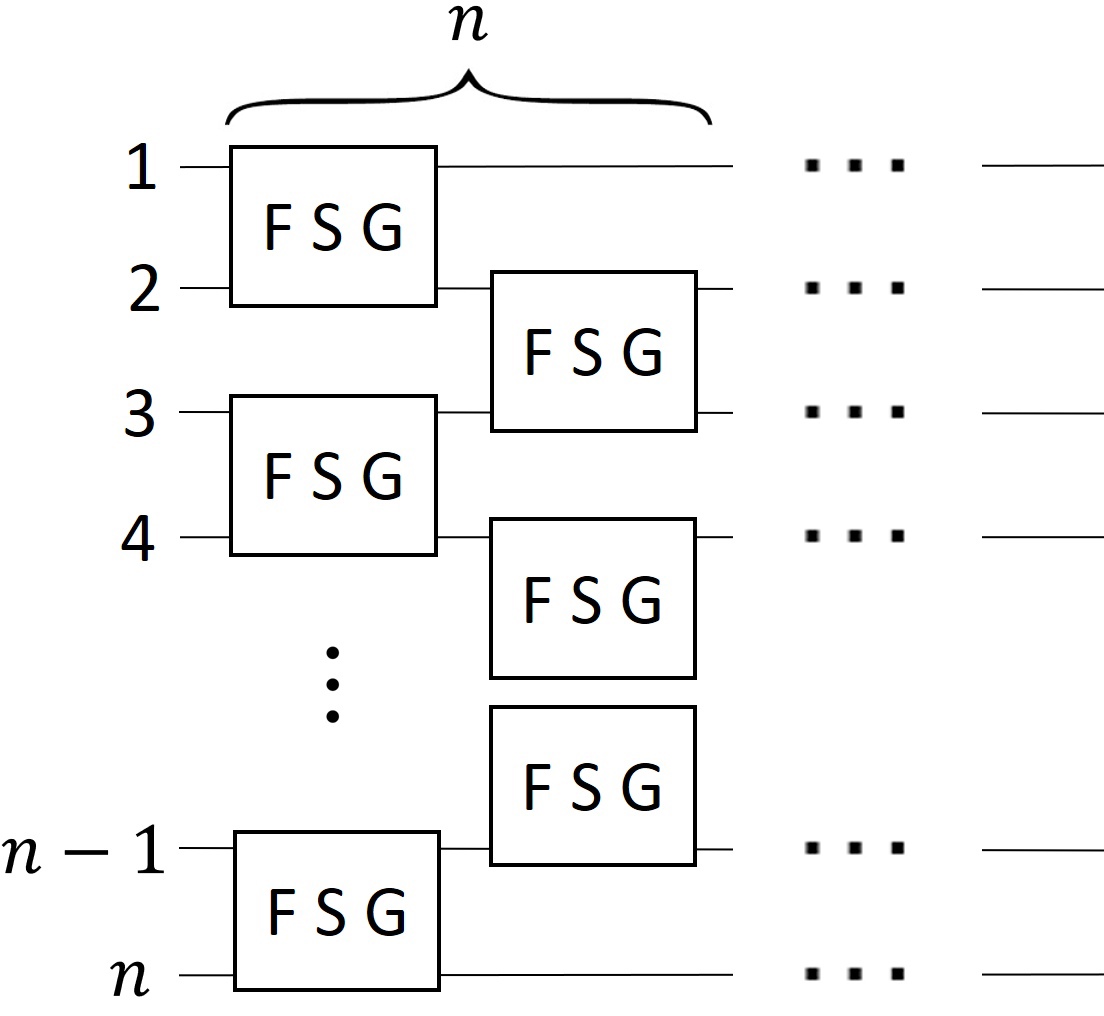}}
\caption{ Quantum circuit for a single Trotter step of the Hamiltonian
$H^\prime$ for an even number of qubits, as a result of which the nodes have the reverse order. }
\label{FSGlay}
\end{figure}

The resulting quantum scheme for a single Trotter step is
represented by layers of FSGs, as shown in Fig. \ref{FSGlay}. It is worth noting that this quantum
circuit corresponds to the evolution of the Hamiltonian $H^\prime$
for any values of $T_{pq}$, $V_{pq}$. This means that this quantum
circuit is appropriate for systems of any dimension described by the 
Hamiltonian (\ref{Htruncated}), since the parameters responsible for the dimension, as well as for hopping and interaction are $V_{pq}$ and $T_{pq}$. Thus, models of
different dimensions are distinct only by the value of the
parameters in the circuit shown in Fig. \ref{ris:imageQC}. Note that all FSGs in Fig. \ref{CNOT} are shown using the
same quantum representation diagrams, but for each FSG, the
internal parameters $\phi$ and $\theta$ differ, in general. 

Each FSG can be accomplished in a digital-analog quantum device with always-on interaction. 
The quantum circuit thus consists of layers of CNOTs between certain qubits interleaved by layers of single-qubit gates. 
Moreover, each layer of CNOTs can be executed with a digital-analog strategy that maximizes the usage of global entangling gates. 
It is
of importance to note that the duration of the interaction between
qubits in a digital-analog quantum device do not depend on the
parameters $\phi$ and $\theta$, since these parameters influence only
single-qubit rotations. The CNOT layer is implemented via a simultaneous
pairwise interaction during the same dimensionless time
$\frac{\pi}{4}$. 

\subsection{Spin 1/2 fermionic Hamiltonians}

In this Subsection, we consider spin-1/2
fermionic systems described by the Hamiltonian
\begin{eqnarray}
\begin{gathered}
H=\sum_{n\neq m}T_{nm}a^{\dagger}_{n\uparrow}a_{m\uparrow}+\sum_{n\neq m}U_{nm}a^{\dagger}_{n\downarrow}a_{m\downarrow}+\\
\sum_n V_{n}a^{\dagger}_{n\uparrow}a_{n\uparrow}a_{n\downarrow}^{\dagger}a_{n\downarrow}.
\end{gathered}
\label{HHubbard}
\end{eqnarray}
 Antisymmetrization must be taken into account only for fermions within each sort. Each qubit of the
chip corresponds to the fermions of one type. Thus, the
number of qubits required to simulate evolution is twice as large as the number of fermionic sites.

\begin{figure}[ht]
\center{\includegraphics[width=0.65\linewidth]{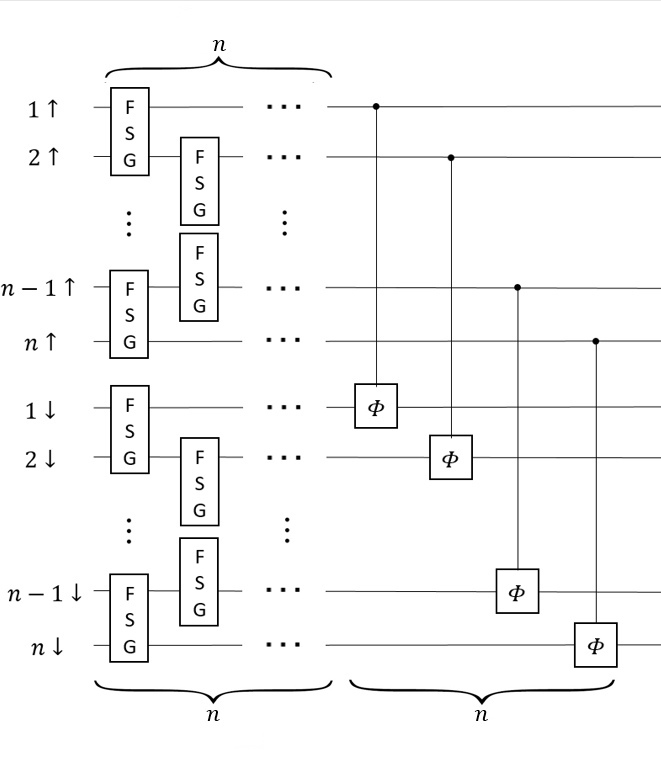} }
\caption{ Quantum circuit for a single Trotter
step of the Hubbard Hamiltonian with two electron spins for an even
number of fermionic nodes.} \label{layers Habbard}
\end{figure}

Consider the term of the Hamiltonian $\sum_{ n\neq m
}T_{nm}a^{\dagger}_{n\uparrow}a_{m\uparrow}$.  The evolution operator
for this contribution, as described above, can be simulated by
layers of FSG operators with $V_{nm}=0$. For the part of the
Hamiltonian $\sum_{n\neq m
}U_{nm}a^{\dagger}_{n\downarrow}a_{m\downarrow}$, the above statement
is also correct. Note that these contributions to the Hamiltonian
belong to different types of fermions, therefore, due to the
commutation relations, they can be simulated independently, as shown
in Fig. \ref{layers Habbard}: the FSG layers
for different types of fermions do not intersect. Thus, for the part of the Hamiltonian (\ref{HHubbard}) $\sum_{n\neq m}T_{nm}a^{\dagger}_{n\uparrow}a_{m\uparrow}+\sum_{n\neq m}U_{nm}a^{\dagger}_{n\downarrow}a_{m\downarrow}$ is simulated using FSG, and the evolution of the part $\sum_n V_{n}a^{\dagger}_{n\uparrow}a_{n\uparrow}a_{n\downarrow}^{\dagger}a_{n\downarrow}$ is implemented using the Cphase gates and the Jordan-Wigner transformation.
Indeed, the part of the Hamiltonian
$\sum_nV_{n}a^{\dagger}_{n\uparrow}a_{n\uparrow}a_{n\downarrow}^{\dagger}a_{n\downarrow}$
under the Jordan-Wigner transformation takes the form
$\sum_n V_{n}\sigma^+_{n\uparrow}\sigma^-_{n\uparrow}\sigma_{n\downarrow}^+\sigma^-_{n\downarrow}$. The Cphase operator, up to the total phase,
is described by the operator $e^{i\phi(Z_\uparrow +Z_\downarrow
-Z_\uparrow Z_\downarrow)}$. Evolution operator for
$V_{n}\sigma^+_{n\uparrow}\sigma^-_{n\uparrow}\sigma_{n\downarrow}^+\sigma^-_{n\downarrow}$
when decomposing the operators $\sigma^\pm=\frac{X\pm iY}{2}$, has
the form $e^{iV_n(Z_\uparrow +Z_\downarrow
-Z_\uparrow Z_\downarrow)t/\hbar}$. Thus, the Cphase operator in Fig. \ref{layers Habbard} is used to
simulate the evolution under the action of
$V_{n}\sigma^+_{n\uparrow}\sigma^-_{n\uparrow}\sigma_{n\downarrow}^+\sigma^-_{n\downarrow}$ with $\Phi=V_nt/\hbar$.

Note that again the parameters of the Cphase and FSG operators are
determined by the coefficients $V_n$, $T_{nm}$, and $U_{nm}$ and are
not equal to each other. For example, the parameter of the Cphase
operator acting on the qubits $1\uparrow$ and $1\downarrow$ is
determined by the coefficient $V_1$ of the Hamiltonian (\ref{HHubbard}), since the FSG operator also
performs a fermionic SWAP operation.

\begin{figure}[ht]
\centering{
\includegraphics[width=0.4\linewidth]{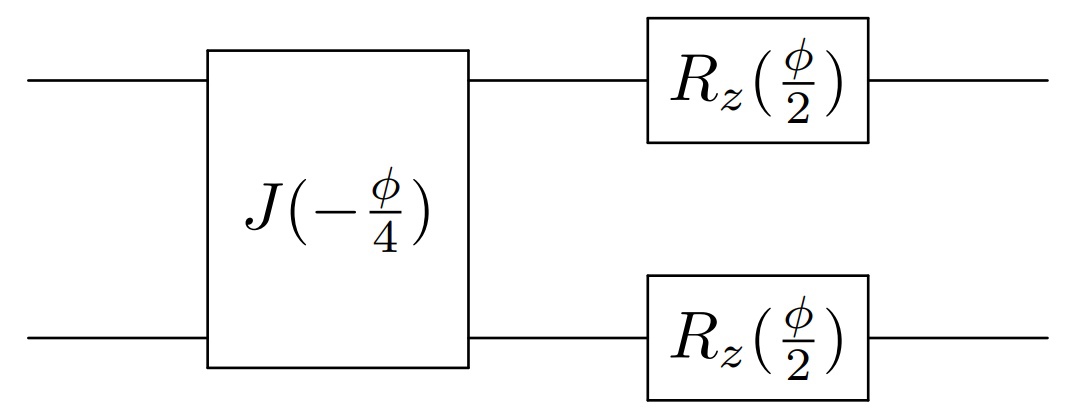}
}
\caption{Implementation  of
Cphase($\phi$) gate on a digital-analog quantum device. When constructing this
circuit, it is assumed that the interaction with other
qubits is disabled. The interaction time cannot be negative, but
adding $2\pi$ to $t$ does not change the result. Hence the
interaction time can always be made positive. } \label{Cphase}
\end{figure}

The difference 
in rotation angles in Cphase gates leads to the difference in interaction times needed to implement 
such operations using a digital-analog strategy. This
complicates the construction of the refocusing operation. In order to 
describe the needed algorithm for this case, we have first to
present a similar algorithm for a set of CNOTs between chosen qubits. 
So, we will discuss the construction of the quantum circuit for the simultaneous implementation of the Cphase gates in Appendix D.

\section{Matrix H(m) as a generator of sequences of states}

The matrix $H(m)$ provides instruction on implementing the refocusing operation. Let us consider two qubits $p$ and $q$ that follow different sequences of states corresponding to the refocusing operation having $m$ time intervals. Also, consider the two columns $A_p$ and $A_q$ of the matrix $A=H(m)$ that define these two distinct sequences. The elements of columns $A_p$ and $A_q$ can be considered as elements of two orthogonal vectors 
\begin{eqnarray}
\sum^{2^m}_{n=1}a_{np}a_{nq}=0;\qquad \forall u,v \quad \vert a_{uv}\vert=1,
\label{elementsA}
\end{eqnarray}
where $a_{np}$ is the $n$ element of the vector $A_p$. The pairwise orthogonality of the vectors $A_p$ follows from the properties of the matrix of the Hadamard operator $H(m)$.

The action of $Z$ on a qubit in the odd state is equivalent to the action of $-Z$ on a qubit in the even state: $Z=-XZX$. Then switching the qubit $i$ to the odd state can be interpreted as changing the sign in the exponent in all the operators $e^{- iZ_pZ_q}$ during the subsequent application of the operator $J$, where $q$ is an index of other qubits participating in the refocusing operation. Since all operators $J_{pq}$ of $J$ commute, the action of the entire refocusing operation is equivalent to the action of the unitary operator $e^B$, where the operator $B$ is the sum of all exponents of the operators $J_{pq}$ participating in the refocusing operation:
\begin{eqnarray}
B=-\frac{i}{2}\sum_{p\neq q; i}b_{pqi}Z_pZ_qt_i,
\label{B} 
\end{eqnarray}
where $b_{pqi}$ is determined by the state of the qubits $p$ and $q$ (odd or even) on the interval $t_i$ of the refocusing operation. Namely, the individual terms of this sum $B$ are the operators $ iZ_pZ_q$ if one of the qubits is in a odd state ($b_{pqi}=-1$), or $ - iZ_pZ_q$ ($b_{pqi}=1$) for all other cases.  Now consider only the connection between qubits $p$ and $q$. That is, consider the contribution to the operator $B$ defined as
\begin{eqnarray}
B_{pq}=-\frac{i}{2}\sum_{i}b_{pqi}Z_pZ_qt_i.
\label{rowB} 
\end{eqnarray}
Suppose that the qubits $p$ and $q$ follow different sequences of states that define the vectors $A_p$ and $A_q$. 
Then $b_{pqi}=a_{ip}a_{iq}$, which implies that
\begin{eqnarray}
\begin{gathered}
B_{pq}=-\frac{i}{2}\sum_{i}a_{ip}a_{iq}Z_pZ_qt_i=\\-\frac{iT}{2^{m+1}}Z_pZ_q\sum_{i=1}^{2^m}a_{ip}a_{iq}=0,
\label{rowBuseA} 
\end{gathered}
\end{eqnarray}
where $T$ is the total refocusing operation time. Equation (\ref{rowBuseA}) is valid, because $t_i=t_j$ for any $i$, $j$.
The number of $t_i$ is even, coinciding with the column length of this matrix. As a result, the interaction during the above period will be compensated during all remaining time intervals $t_i$. Thus, since any two vectors formed by the columns of the matrix $H(m)$ are orthogonal, the interaction between all the qubits that follow different sequences is compensated. However, if two qubits follow the same sequence, they will be affected by the desired operator $e^{- iZ_pZ_qT}$, where $T$ is the total time of the refocusing operation, $p$ and $q$ are the indices of qubits that follow the same sequence.

\section{Number of sequences for an arbitrary connectivity topology}

We have previously defined in  Eq. (\ref{condition}) the minimum number of state sequences for the all-to-all connectivity topology. 
This Appendix shows how to determine this number for the general case.

First, let us construct the refocusing operation, which is equivalent to the action of the identity operator $I$. In this case, no interaction pair exists. This example is fundamental because it forms a basis for constructing any refocusing operation. Consider the all-to-all connectivity topology of $k=m$ qubits. For implementing the refocusing operation, it is necessary to have $m$ sequences of states, according to Eq. (\ref{condition}). If we add an extra qubit to the network, i.e., the qubit $k+1$, which has a connection to all qubits except for the $1$'st qubit, then the number of sequences of states for implementing the refocusing operation does not change. This is due to the fact that the $k+1$'th qubit can repeat the parity states of the $1$'st qubit during all time intervals of the refocusing operation. Let us add one extra qubit, which again has connections to all qubits except the first one. To construct a refocusing operation, it is already necessary to have $m+1$ sequences of states, since the qubits from $2$ to $k+2$ form an all-to-all topology of $k+1$ qubits, according to Eq. (\ref{condition}). From this analysis, we draw an important conclusion that adding a qubit that does not change the size of the largest graph with all-to-all connectivity does not increase the number of necessary sequences of states.

Let us now consider a network of $k$ qubits with an arbitrary connectivity topology. Without reducing the generality, we assume that the qubits from $1$ to $k^\prime$ form the largest all-to-all connectivity cluster. In order to find the necessary number of sequences, we first consider this cluster and then start adding the remaining qubits one by one according to the given connectivity topology. It directly follows from our analysis that the number of sequences does not change after each step. Thus, in the case of an arbitrary connectivity topology, the required number of sequences of states is determined by the size of the largest all-to-all connectivity cluster that it contains.

\begin{figure}[ht]
\center{\includegraphics[width=0.5\linewidth]{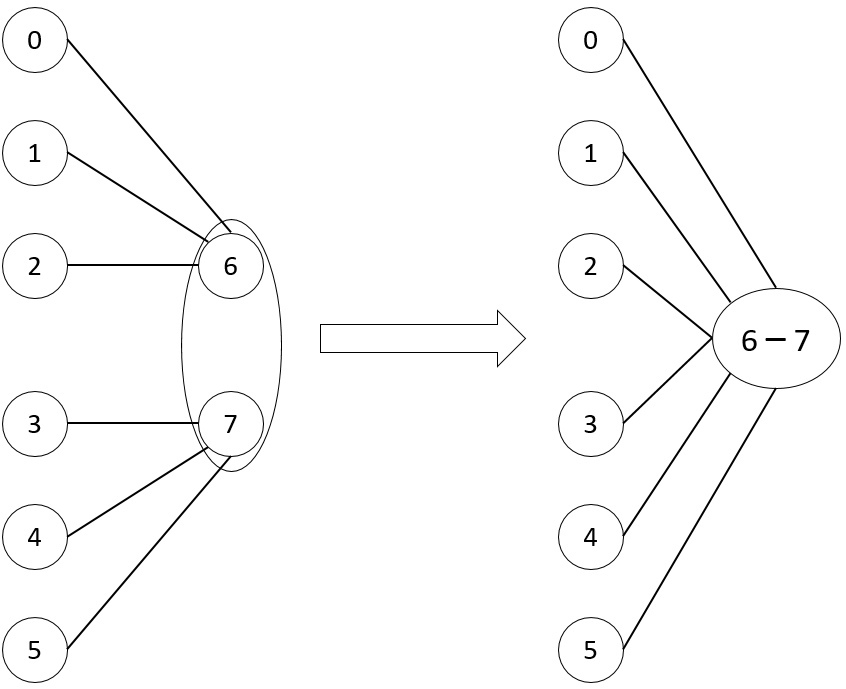} }
\caption{Layout of the deformation of the connectivity topology due to the interaction pair formed by the qubits $6$ and $7$. For clarity, this layout does not show the scheme of interaction between the qubits $0$ -- $5$, since it does not change.}
\label{deformation}
\end{figure}

Next, we must consider interaction pairs. Since the qubits forming the interaction pair follow the same sequence of states, each qubit from this pair must have a different sequence of states, with all qubits interacting with the qubits of the pair. The latter fact can be understood in terms of the deformation of the initial topology: qubits belonging to the same pair must be merged. Such a new node (a combined pair) has connections with all qubits with which at least one qubit forming an interaction pair had a connection. The layout of the deformation process is shown in Fig. \ref{deformation}. As a result of the deformation, a new connectivity topology is formed. Determining the required number of sequences of states for such systems without interaction pairs has already been demonstrated above (a refocusing operation which is equivalent to the action of the identity operator $I$). Thus we explained how to find a required number of sequences of states to construct the refocusing operation for the arbitrary connectivity topology.

In general, the process of simultaneous implementation of local entangling gates (CNOTs, SWAPs, Cz, Cphase) consists of the following steps: 

(i) identification of interaction pairs depending on the general algorithm we wish to implement, 

(ii) determination of the required number of sequences of states according to the general rules of Appendix B (matrix $H(m)$), which depends on the connectivity (dimensionality) of the problem, 

(iii) construction of a quantum scheme implementing a refocusing operation based on $H(m)$.

\section{Spin-1/2 fermionic Hamiltonian: Cphase gates with always-on interaction}

This Appendix demonstrates the possibility of implementing the Trotter step for the Fermi-Hubbard Hamiltonian with spin $1/2$. According to the quantum circuit shown in Fig. \ref{layers Habbard}, for that, it is necessary to implement Cphase gates as well as FSG layers. In Fig. \ref{layers Habbard}, the parameters of the Cphase gate are determined by the coefficients $V_n$ in the Hamiltonian (\ref{HHubbard}), which in general differ from each other. Hence for each interaction pair, the required duration of the action of $H_{int}^{p,q}$, in the general case, is different according to the circuit shown in Fig. \ref{layers Habbard}. Note that this condition is similar to the requirement in the previously described case of systems with the spread of interaction parameters (Section IV), where each interaction pair existed for a different time. In the case of the Cphase implementation, for each interaction pair formed by the qubits that the Cphase gate acts on, $H_{int}^{p,q}$ acts for a different time. However, the previously described approach is also applicable in this case.

The parameters of each Cphase gate are determined by the Hamiltonian (\ref{HHubbard}). Hence, there is a Cphase operator for which the required interaction time is maximum. Similar to the case of the chip with the spread of interaction parameters, the total time of simultaneous implementation of all Cphase gates is exactly this maximum time. 
The interaction pair with the minimum interaction constant exists during all time intervals. Interaction pairs for other Cphase gates require less time, so there is an excess time for each of them. During this time, the interaction pair is destroyed because the qubits form and follow different states' sequences.

All the arguments described above are correct in the case of simultaneous implementation of Cphase gates on a digital-analog quantum device with the spread of interaction parameters. However, for each interaction pair, the duration of interaction depends not only on the coefficients of the Hamiltonian (\ref{HHubbard}) but also on the parameter $\alpha_{pq}$, where $p$ and $q$ are the qubits forming the interaction pair.

\begin{figure}[ht]

\center{\includegraphics[width=0.35\linewidth]{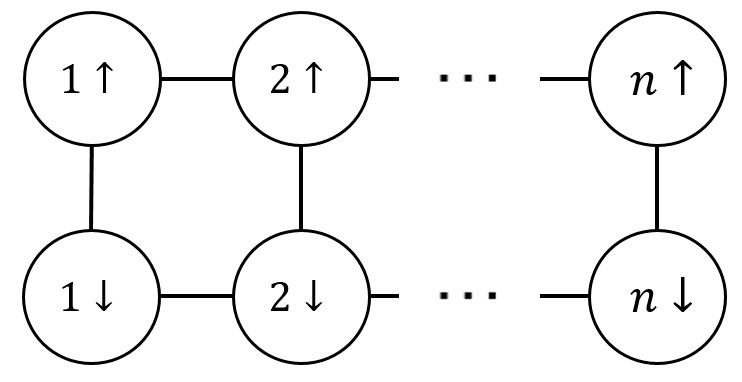} }

\caption{ Two-dimensional qubit connectivity scheme for the quantum simulation of spin-1/2 Fermi-Hubbard model in any dimension.
}
\label{idealHabbardlayout}
\end{figure}

Let us consider the connectivity scheme shown in Fig. \ref{idealHabbardlayout}. Such a topology is most suitable for simulating the evolution of the system described by the Hamiltonian (\ref{HHubbard}) according to the quantum scheme shown in Fig. \ref{layers Habbard}. To implement the algorithm described above, two sequences determined by the matrix $H(2)$ are sufficient. This result is similar to the case of the odd step shown in Fig. \ref{Trottertwodimensiona}, where the interaction pair had a staggered pattern order.


We now estimate the complexity of the implementation of the quantum circuit shown in Fig. \ref{layers Habbard} in view of the connectivity scheme shown in Fig. \ref{idealHabbardlayout}. In the case of digital-analog quantum devices with identical interaction parameters, the Cphase gates implementation requires $\mathcal{O}(n)$ multi-qubit operations. Implementing FSGs layers also requires $\mathcal{O}(n)$ multi-qubit operations. Hence the circuit depth is $\mathcal{O}(n)$ gates. In the case of a chip with the spread in the interaction parameters, the complexity of the realization of Cphase gates does not change, as described above. The implementation of FSGs layers requires $\mathcal{O}(n^2)$ multi-qubit gates. Hence the circuit depth, in this case, is $\mathcal{O}(n^2)$.

In a particular case, when all coefficients $V_n$ are the same, there appears a significant simplification in the case of a chip with identical interaction parameters. The equality of all the coefficients of $V_n$ implies the same desired interaction time within each interaction pair. Thus, the number of multi-quit operations for implementing Cphase gates grows as $\mathcal{O}(1)$.


\end{document}